\documentclass[twocolumn, trackchanges]{aastex63}

\newcommand{\edenAP}{\texttt{edenAP}}
\newcommand{\TLS}{\texttt{TLS}}
\newcommand{\TESS}{\textit{TESS}}
\newcommand{\Spitzer}{\textit{Spitzer}}
\newcommand{\ktwo}{\textit{K2}}

\usepackage{array}
\usepackage{hyperref}
\usepackage{longtable}
\usepackage{rotating, graphicx}
\newcolumntype{P}[1]{>{\centering\arraybackslash}p{#1}}

\accepted{February 20, 2020}
\submitjournal{AJ}

\shorttitle{Detection Limits for Nearby Late Red Dwarfs}
\shortauthors{Gibbs et al.}

\begin{document}

\title{EDEN: Sensitivity Analysis and Transiting Planet Detection Limits for Nearby Late Red Dwarfs}



\author{Aidan Gibbs}
\affiliation{Steward Observatory, The University of Arizona, 933 N. Cherry
Avenue, Tucson, AZ 85721, USA}
\affiliation{Department of Physics \& Astronomy, University of California, Los Angeles, Los Angeles, CA 90095, USA}

\author{Alex Bixel}
\affiliation{Steward Observatory, The University of Arizona, 933 N. Cherry
Avenue, Tucson, AZ 85721, USA}
\affiliation{NASA Nexus For Exoplanetary System Science: Earths in Other Solar Systems Team}

\author{Benjamin V. Rackham}
\affiliation{Department of Earth, Atmospheric, and Planetary Sciences and Kavli Institute for Astrophysics and Space Research, Massachusetts Institute of Technology, 77 Massachusetts Ave., Cambridge, MA 02139, USA}
\altaffiliation{51 Pegasi b Postdoctoral Fellow}

\author{D\'aniel Apai}
\affiliation{Steward Observatory, The University of Arizona, 933 N. Cherry
Avenue, Tucson, AZ 85721, USA}
\affiliation{NASA Nexus For Exoplanetary System Science: Earths in Other Solar Systems Team}
\affiliation{Lunar and Planetary Laboratory, The University of Arizona, 1640 E. University Boulevard,
Tucson, AZ 85718, USA}

\author{Martin Schlecker}
\affiliation{Max Planck Institute for Astronomy, Konigstuhl 17, D-69117 Heidelberg, Germany}

\author{N\'estor Espinoza}
\affiliation{Space Telescope Science Institute, 3700 San Martin Drive, Baltimore, MD 21218, USA}
\altaffiliation{IAU-Gruber Fellow}

\author{Luigi Mancini}
\affiliation{Department of Physics, University of Rome Tor Vergata, Via della Ricerca
Scientifica 1, I-00133 Rome, Italy}
\affiliation{Max Planck Institute for Astronomy, Konigstuhl 17, D-69117 Heidelberg, Germany}
\affiliation{INAF – Osservatorio Astrofisico di Torino, via Osservatorio 20, I-10025
Pino Torinese, Italy}
\affiliation{International Institute for Advanced Scientific Studies (IIASS), Via G.
Pellegrino 19, I-84019 Vietri sul Mare (SA), Italy}

\author{Wen-Ping Chen}
\affiliation{Graduate Institute of Astronomy, National Central University, 300 Jhongda Road, Zhongli, Taoyuan 32001, Taiwan}

\author{Thomas Henning}
\affiliation{Max Planck Institute for Astronomy, Konigstuhl 17, D-69117 Heidelberg, Germany}

\author{Paul Gabor}
\author{Richard Boyle}
\affiliation{Vatican Observatory Research Group, University of Arizona, 933 N Cherry Ave., Tucson AZ, 85721-0065, USA}

\author{Jose Perez Chavez}
\author{Allie Mousseau}
\author{Jeremy Dietrich}
\author{Quentin Jay Socia}
\affiliation{Steward Observatory, The University of Arizona, 933 N. Cherry
Avenue, Tucson, AZ 85721, USA}

\author{Wing Ip}
\author{Chow-Choong Ngeow}
\author{Anli Tsai}
\affiliation{Graduate Institute of Astronomy, National Central University, 300 Jhongda Road, Zhongli, Taoyuan 32001, Taiwan}

\author{Asmita Bhandare}
\author{Victor Marian}
\author{Hans Baehr}
\author{Samantha Brown}
\author{Maximilian H\"aberle}
\author{Miriam Keppler}
\author{Karan Molaverdikhani}
\author{Paula Sarkis}
\affiliation{Max Planck Institute for Astronomy, Konigstuhl 17, D-69117 Heidelberg, Germany}

\begin{abstract}

Small planets are common around late-M dwarfs and can be detected through highly precise photometry by the transit method. Planets orbiting nearby stars are particularly important as they are often the best-suited for future follow-up studies. 
We present observations of three nearby M-dwarfs referred to as EIC-1, EIC-2, and EIC-3, and use them to search for transits and set limits on the presence of planets. On most nights our observations are sensitive to Earth-sized transiting planets, and photometric precision is similar to or better than {\TESS} for faint late-M dwarfs of the same magnitude ($I\approx 15$ mag). 
We present our photometry and transit search pipeline, which utilizes simple median detrending in combination with transit least squares based transit detection \citep{Hippke2019}. For these targets, and transiting planets between one and two Earth radii, we achieve an average transit detection probability of${\sim} 60\%$ between periods of 0.5 and 2 days, ${\sim} 30\%$ between 2 and 5 days, and ${\sim} 10\%$ between 5 and 10 days. These sensitivities are conservative compared to visual searches.

\end{abstract}



\section{Introduction} \label{sec:intro}

 Planetary systems around nearby stars are set to play a particularly important role in the future of exoplanet characterization studies, yet only a very small fraction of these planets have been identified to date. Reconnaissance spectroscopy of nearby, small (Earth-sized) transiting planets is possible now with the \textit{Hubble Space Telescope} (e.g., as in the TRAPPIST-1 system, see \citealt{deWit2016,deWit2018,Zhang2018,Wakeford2019}) and in-depth spectroscopic studies of these systems will be possible in the near-future with the \textit{James Webb Space Telescope} \citep[e.g.,][]{Greene2016,Morley2017,Lustig-Yaeger2019} and with the \textit{ARIEL} mission \citep[e.g.,][]{Tinetti2018}. Transiting, habitable-zone, Earth-sized planets around nearby stars are likely to be the only type of habitable planets that can be characterized in detail in the next two decades.

Although only a fraction of planets happen to transit as observed from Earth, fortunately, the high frequency of M-dwarfs in the solar neighborhood, the most favorable host stars for detecting Earth-sized planets, improves the chances of a positive detection. Based on results from the RECONS group \citep[][]{Henry2018}, there are 283 currently known M-type stars within 10 pc, and that number continues to grow. In addition, small (1--4\,$R_\Earth$) planets are found to be very common around M-dwarfs \citep{DressingCharbonneau2015,Mulders2015a,Mulders2015b,HardegreeUllman2019}. However,  M-dwarfs in the solar neighborhood are located isotropically in the sky, requiring targeted, star-by-star monitoring (e.g., \citealt{Nutzman2008,Jehin2011,Delrez2018}). Worldwide networks of ground-based telescopes that can obtain continuous targeted coverage are therefore well-suited to search for these planets \citep{Blake2008}. 

The Exoearth Discovery \& Exploration Network (EDEN, PIs: D. Apai, P. Gabor, Th. Henning, W-P. Chen) is a multi-continental research network that searches for habitable-zone planets within fifty lightyears\footnote{http://project-eden.space}. EDEN's transit survey component began in Spring 2018 and currently uses eight telescopes to search for transiting planets around nearby late M-dwarf stars, which are the easiest stars to find Earth-sized planets around. EDEN differs from other ongoing surveys in that it uses several large preexisting telescopes ($>$1-m diameter) and that its longitudinally distributed stations are capable of providing continuous coverage.
When no planet is found in a system, EDEN also aims to place stringent upper limits on the probability that short-period planets are present. The interpretation of such non-detections requires a robust and consistent observing strategy, thorough understanding and modeling of systematics, efficient photometric pipeline and trend removal (detrending), and a well-characterized planet-detection algorithm. 
With this photometric and detection pipeline, EDEN also provides an excellent telescope network for photometric follow-up of planet candidates identified by NASA's {\TESS} \citep{RickerTESS2015} transit search mission.

We review here these components of our sensitivity analysis, and present example results for the first three EDEN targets searched in depth. We do not detect any convincing transit candidates for follow-up, but show that there is a high probability we would have detected Earth-sized planets with periods less than 5 days if their orbital planes were aligned with our line of sight. In Section~\ref{sec:obs} we briefly describe the EDEN telescopes and our observational methods. Section~\ref{sec:data} details our data reduction pipeline before lightcurve detrending and transit search described in Section~\ref{sec:search}. In Section~\ref{sec:targets}, we provide background on the selected EDEN targets for which we perform a sensitivity analysis in Section~\ref{sec:limits}. Finally, in Section~\ref{sec:discussion} we discuss our planet detection limits in the context of M-dwarf planetary occurence rates, known systems, and NASA's \TESS mission.

\section{Observations} \label{sec:obs}

We briefly describe the EDEN telescopes, survey target selection, and photometric data collection procedures in order to provide context for our data reduction, transit search, and sensitivity analysis methods. A nuanced discussion of our strategy for selecting and observing targets, and a comparison with other surveys, will be reserved for a future paper (Apai. et al, in prep.), and only necessary details are included here. 

\subsection{Observatories} \label{sec:teles}

EDEN observations are currently conducted with eight unique telescopes at seven observatories in North America, Europe, and Asia. The telescopes are the Kuiper $1.55$\,m (Mount Bigelow, Arizona), Bok $2.3$\,m (Kitt Peak, Arizona), Vatican Advanced Technology Telescope $1.8$\,m (VATT; Mount Graham, Arizona), Phillips $0.6$\,m and Schulman $0.8$\,m (Mount Lemmon, Arizona), Calar Alto $1.23$\,m (Calar Alto, Spain), Cassini $1.52$\,m (Mount Orzale, Italy), and Lulin $1$\,m (Mount Lulin, Taiwan). Table \ref{tab:teles} details the location, design, and CCD imager of each telescope. With the exception of the robotic Schulman and Phillips telescopes, each of them is manually controlled by an observer, who actively monitors weather conditions and instrument performance during the course of a night. While the telescope designs are varied, 
each of the telescopes has been carefully evaluated for photometric performance before its inclusion in EDEN and, when necessary, changes have been made in the telescope's operation and setup, which will be detailed in Apai et al. (in prep.). Systematic differences between telescopes therefore have very minor effects on the final lightcurves and transit search. These differences can be compensated for during the data reduction and detrending steps, discussed in Section~\ref{sec:data} and \ref{sec:search}. 

The majority of the EDEN telescopes are not solely dedicated to EDEN, so observations are scheduled at each facility individually in blocks usually from two to ten days per month, depending on availability. Observing science targets at these sites has been ongoing since June 2018 (following a six-month-long EDEN pilot program), with observations of the targets discussed in this paper occurring between June 2018 and February 2019.

\begin{deluxetable*}{P{1cm}P{1.5cm}P{1cm}P{1cm}P{1.5cm}P{1cm}P{1.5cm}P{1cm}P{1.5cm}}[t]
\tablecaption{EDEN Telescopes \label{tab:teles}}

\tablehead{\colhead{Telescope} & \colhead{Location} & \colhead{Operation} & \colhead{Mount} & \colhead{CCD Imager} & \colhead{Det. Size} & \colhead{FOV} & \colhead{Px. Scale} & \colhead{$Q_e$ at 700 nm}} 

\startdata
Phillips $0.6$\,m &  Mount Lemmon, Arizona & Robotic & EQ  & SBIG STX (KAF-16803) & 4096$\times$4096 & $22'{\times}22'$ & 0.35" & $40\%$  \\
\hline
Schulman $0.8$\,m &  Mount Lemmon, Arizona & Robotic & EQ & SBIG STX (KAF-16803) & 4096$\times$4096 & $22'{\times}22'$ & 0.35" & $40\%$\\
\hline
Lulin $1.0$\,m &  Mount Lulin, Taiwan & Classical & EQ & Sophia 2048B CCD & 2048$\times$2048 & $13.08'{\times}13.08'$ & 0.39" & 60\% \\
\hline
Calar Alto $1.23$\,m &  Calar Alto, Spain & Remote & EQ & DLR-MKIII camera with e2v CCD231-84-NIMO-BI-DD sensor & 4k$\times$4k & $21.5' \times 21.5'$ & 0.31" & 93\%\\
\hline
Cassini $1.52$\,m &  Mount Orzale, Italy & Classical & EQ & Bologna Faint Object Spectrograph and Camera & 1300$\times$1340 & $13'{\times}12.6'$ & 0.34" & 75\% \\
\hline
Kuiper $1.55$\,m &  Mount Bigelow, Arizona & Classical & EQ & Mont4K SN3088 \citep{Weiner2018} & 4096$\times$4097 & $9.7'{\times}9.7'$ & 0.14" & 62\%  \\
\hline
VATT $1.8$\,m &  Mount Graham, Arizona & Classical & Alt-Az & VATT4K STA0500A CCD & 4064$\times$4064 & $12.5'{\times}12.5'$ & 0.188" & 80\%  \\
\hline
Bok $2.3$\,m & Kitt Peak, Arizona & Classical & EQ & 90 Prime Focus Wide-Field Imager \citep{Williams2004} & 4$\times$4032$\times$ 4096 & $1.16^\circ {\times} 1.16^\circ$ & 0.4" & $80\%$  \\
\enddata



\end{deluxetable*}

\subsection{Target Selection} \label{sec:targ_select}

EDEN's primary focus is to search for potentially habitable planets within 15 pc ($\sim$50 lightyears). Correspondingly, for the EDEN Transit Survey, our target selection prioritizes M4 and later-spectral-type host stars, which offer favorable planet-to-star projected areal ratios, making broadly Earth-sized planets detectable in our data. We eliminate known close binary stars that may reduce the stability of putative planets and would complicate the interpretation of the lightcurve. We then prioritize sources that are too faint (I$>$15 mag) to be efficiently searched by TESS or are outside TESS's sky coverage. In addition to these high-priority EDEN targets we also include separately targets of particular interest in our source catalog. Such targets may be exoplanet candidate host stars (from radial velocity or transit searches), for which EDEN data can prove valuable for candidate verification. Such follow-up targets (where prior knowledge about a planet’s presence exists) will not be used in future exoplanet occurrence rate studies.

\subsection{Science Observations} \label{sec:sci}

EDEN targets, including those discussed in this paper, are late-M dwarfs scattered throughout the Northern Hemisphere sky and thus must be observed one at a time. For planets orbiting within or interior to the habitable zone of these stars \citep[e.g.,][]{Kasting1993, Kopparapu2014}, expected transit durations range from $0.5$ to $3$\,hours at periods of roughly $0.5$ to $10$\,days. 

To maximize the probability of observing transits with these parameters and to take advantage of the longitudinal coverage of EDEN telescopes, we designed our observing strategy around two pillars. First, we observe each target for as long as possible on a given night. This typically means that on a clear night we observe a primary target for $>$6\,hr, and then a secondary target for 2--3\,hr when the primary is not observable. This also increases the chance of observing a full transit, which is easier to detrend and detect than fractional transits. Second, whenever possible, we schedule simultaneous observing campaigns in Arizona, Europe, and Taiwan to allow the potential for continuous $24$\,hour monitoring of one target for multiple days. On such longer, coordinated runs---given good weather at all sites---we can obtain roughly week-long continuous sequences, limited only by our allocated time on these facilities.

These pillars allow us to quickly get good phase coverage of a target for shorter-period planets. Practically, continuous observation has been difficult to fully exploit because of the rarity of getting good weather on three continents during the entire run. The number of nights dedicated to any target is based on the probability that we would have observed two transits of a planet with an orbital period of less than 10 days. As this probability increases, we deprioritize a given target so that more targets can be adequately sampled. While it is not practically possible to reach $100\%$ detection probability for planets throughout the entire habitable zone (from inner to outer edge, \citealt{Kopparapu2014}), we aim to reach high sensitivity for transiting planets that orbit at the inner edge of the habitable zone (i.e., ${\sim} 50\%$ successful detection of Earth-size transiting planets), which typically translates to some sensitivity (${\gtrsim} 10\%$) throughout the habitable zone.

\subsubsection{Observational Procedures} \label{sec:procedures}

Although each of our telescopes has somewhat different capabilities and performance, we adopt the same observational procedures at each telescope to minimize systematic differences.

\paragraph{Filter} For each telescope we use a near-infrared (NIR) (or blue-blocking) filter, such as Harris-I or similar. This filter choice maximizes the collected photons from our targets, which are brightest in the NIR, while blocking unwanted sky background from the Moon and skyglow. Since Spring 2019, the filter has been standardized at all telescopes to an uncoated GG~495\footnote{\href{https://www.us.schott.com/d/corporate/9e67ed08-32ee-422c-b517-5496b0ff7cf3/schott-longpass-gg495-jun-2017-en.pdf}{www.us.schott.com}} glass long-pass filter  (transparent at $> 500$ nm). Redder filters such as \textit{I} or \textit{z'} have been occasionally used for bright targets if the sky background is very high, for example, during a full moon. The \textit{z'} is otherwise generally avoided because of the low quantum efficiency of most CCD detectors at those wavelengths and the greater presence of telluric absorption bands from water vapor \citep{BailerJones2003, Blake2008}.
\paragraph{Exposure Time} The exposure time is chosen to balance competing signal-to-noise and cadence considerations. We never allow the peak target flux to go above ${\sim} 60\%$ the detector's full well, where the detector begins to exhibit non-linear behavior. In a given period of time, such as a transit duration, the total Poisson-noise-driven signal-to-noise ratio (SNR) follows the relationship $$SNR_\mathrm{tot}\propto \sqrt{\frac{R}{1+R}},$$ where $R$ is the ratio of the exposure time to readout time \citep{Howell2019}. This relationship levels off at $R \sim 3.5$, and we thus aim for an exposure time of ${\sim} 3.5\times$ the readout time. For our telescopes with a diameter larger than one meter and targets with magnitude $I \sim 14$, this gives a cadence $<$60\,s.    
\paragraph{Focus} Previous work (e.g., \citealt{Southworth2009}) has shown that defocusing can result in more precise lightcurves as the point spread function (PSF) is spread across more pixels. We aim for a slight-to-moderate defocus of 2--3", so that pixel-to-pixel variations are reduced, but the PSF maintains a Gaussian shape. Since defocusing also reduces the peak of the PSF, it has the additional benefit of allowing longer exposures.

\subsection{Calibration Frames} \label{sec:calibframes}

We follow standard calibration procedures for flat, bias, and dark corrections to reduce systematic effects on our lightcurves. Detailed tests (complete re-reduction and analysis of selected datasets) show that the details of the basic calibration do not affect the resulting lightcurve precision significantly.

For our calibration procedure, before or after every night of observation, we collect ${\sim} 10$ twilight flat-fields with exposure times chosen to maintain a sky flux approximately at $50\%$ the detector's full well, the same as our desired peak target flux. In some cases of inclement weather during twilight, we may use dome flats, but these are not preferred since they have less uniform illumination. The minimum flat exposure time is always long enough so that the shutter time has ${<}1\%$ effect on the precision of the flat.   

Generally, at least once per observing run, we collect a set of bias and dark frames. The dark current for our exposure times is nearly zero at all telescopes and is usually not subtracted. At some telescopes darks are not collected for this reason. There is no evidence for persistence on any of our detectors. 

\section{Data Reduction} \label{sec:data}

EDEN data reduction is performed with a custom Python-based automatic pipeline, {\edenAP}, which is based on a precursor pipeline for reducing Las Cumbres Observatory Global Telescope (LCOGT) lightcurves \citep{Brown2013}. {\edenAP} is designed to accommodate the particularities of the individual telescopes in the EDEN telescope network and reduce the data in a consistent manner. Differences that must be accounted for include number and configuration of chip amplifiers, and pixel scale. {\edenAP} is called locally when new raw data arrive, and produces a comparison-star-detrended (Section \ref{sec:compdetrend}) lightcurve for each observation as its final output, which can be further detrended and used for a transit search. The pipeline is highly automated and, in the event of improvements to the algorithm, {\edenAP} can be re-run on all previous data with minimal effort. All raw data are stored at the University of Arizona, as well as through a cloud storage provider (Amazon Web Services).

\subsection{Science Calibration} \label{sec:scical}

The first step in {\edenAP} is to calibrate the raw science frames using the calibration frames discussed in Section~\ref{sec:calibframes}. In the event that calibration frames are not available or are of poor quality, this step can be skipped with the rest of the pipeline remaining the same. To create master calibration frames, we collect all bias frames within one month of the observation, and all dark and flat frames within the observation run. Monitoring of flat fields has indicated that these stay mostly constant over the course of a run, with the exception of minor localized dust accumulation and chance occurrences such as insects getting trapped in the optical path. In cases where many hundreds of calibration frames are available in the above time periods, we narrow the period and only collect calibration frames within two to three days of the observation.

\subsection{Astrometry} \label{sec:astrometry}

We then derive the astrometric solution for every science frame by using a local installation of the \texttt{astrometry.net} software package \citep{Lang2010}. While this solution provides accurate astrometric calibration for most frames, it can fail in case of partial cloud cover or poor seeing. If no astrometric solution can be found for a particular image, the solution from the preceding image is used, despite these data typically being very poor. The astrometric solution derived is used as a first guess for placing photometric apertures, however, we always refine the centroid using the \texttt{photutils}\footnote{\url{https://photutils.readthedocs.io/en/stable/}} \texttt{DAOStarFinder} method \citep{Bradley2019}, based on the DAOFIND algorithm \citep{Stetson1987}. Position refinement is key to getting sub-pixel centroid precision, especially for our high proper motion target stars.

\subsection{Photometry} \label{sec:phot}

Aperture photometry is performed using the \texttt{photutils} package \citep{Bradley2019}. For every star in the field of view, we measure the intensity in apertures ranging from 5 to 50 pixels in steps of 1 pixel. The aperture size that minimizes the RMS scatter of the target star lightcurve is selected as the best aperture for all sources. The optimal size depends on detector and seeing, but typical size are roughly a few arcseconds. Sky background is calculated as the median of a $60{\times}60$ pixel sub-image around the star with other sources clipped. Photometry is saved into a Python pickle file with other important information for each star, such as centroid positions, stellar magnitudes, background, FWHM, airmass, etc., which can later be used for detrending steps and vetting transit-like signals.

\subsection{Comparison Star Detrending} \label{sec:compdetrend}

The final step in {\edenAP} is to detrend the target lightcurve on the basis of comparison star lightcurves. Trends are long or short term photometric variations in the lightcurve that decrease transit detection sensitivity, and can arise from instrumental, atmospheric, and stellar variability. We select the best comparison stars by first filtering out stars that are saturated, are too faint (several magnitudes dimmer than the target) or have too many failed photometric measurements. Next, we divide the flux normalized target lightcurve by the normalized lightcurves of every comparison star, and rank them based on the average standard deviation in windows of 20 data points. The six with the lowest average deviation (i.e. those with the most similar data trends) are median-combined into a ``super comparison'' lightcurve, which is then divided from the target lightcurve. For crowded fields with many available comparison stars, it is conceivable that this selection method could weaken or remove transit signals. We believe this is highly unlikely, however, due to the improbability that comparison lightcurves would have the necessary shape to remove a transit, and because the duration of the window is shorter than any expected non-grazing transit. Nevertheless, we account for this in our sensitivity analysis (Section \ref{sec:analysis}) by re-selecting comparison stars after injecting transits.

\section{Transit Search} \label{sec:search}

In the subsequent steps we identify and remove residual systematic trends (i.e., those not shared fully by comparison stars) and search for lightcurve features that are candidate transit events.
Our approach is a modular, automatic, step-by-step process that is robust and easily repeatable, allowing for detailed test runs and process optimization. As detailed in the following subsections, we use a simple median-detrending method and base our vetting methods on instrumental parameters, such as airmass and centroid position, to attempt to explain observed trends and transit-like features. The end result is either a promising candidate, triggering follow-up observations, or sensitivity limits if no convincing candidate is found. A discussion of transit candidate follow-up is reserved for a future paper (Apai et al., in prep.).

\subsection{Interactive Data Viewer} \label{sec:viewer}

We visually inspect every lightcurve on a single EDEN target to ensure that lightcurve anomalies are recognized and managed correctly. We select high-quality data for further analysis without relying on automatic algorithms. To streamline this process, we have implemented an interactive data viewer that displays each lightcurve along with systematic trends, allowing the user to flag large sections of problematic data (e.g., stellar flares, passing clouds) for removal and points of interest (a transit-like feature) for further analysis. Excluding poor-quality data is exceedingly important because strong systematic trends can be fit as transits, and they can throw off the correct period determination if one transit of an otherwise detectable period happened to occur within it. 
Individual outlier data points are ignored in this step, but are efficiently removed by our automatic filtering in the next step.

\subsection{Median Detrending} \label{sec:meddetrend}

\begin{figure*}[t]
\centering
\includegraphics[width=0.9\textwidth]{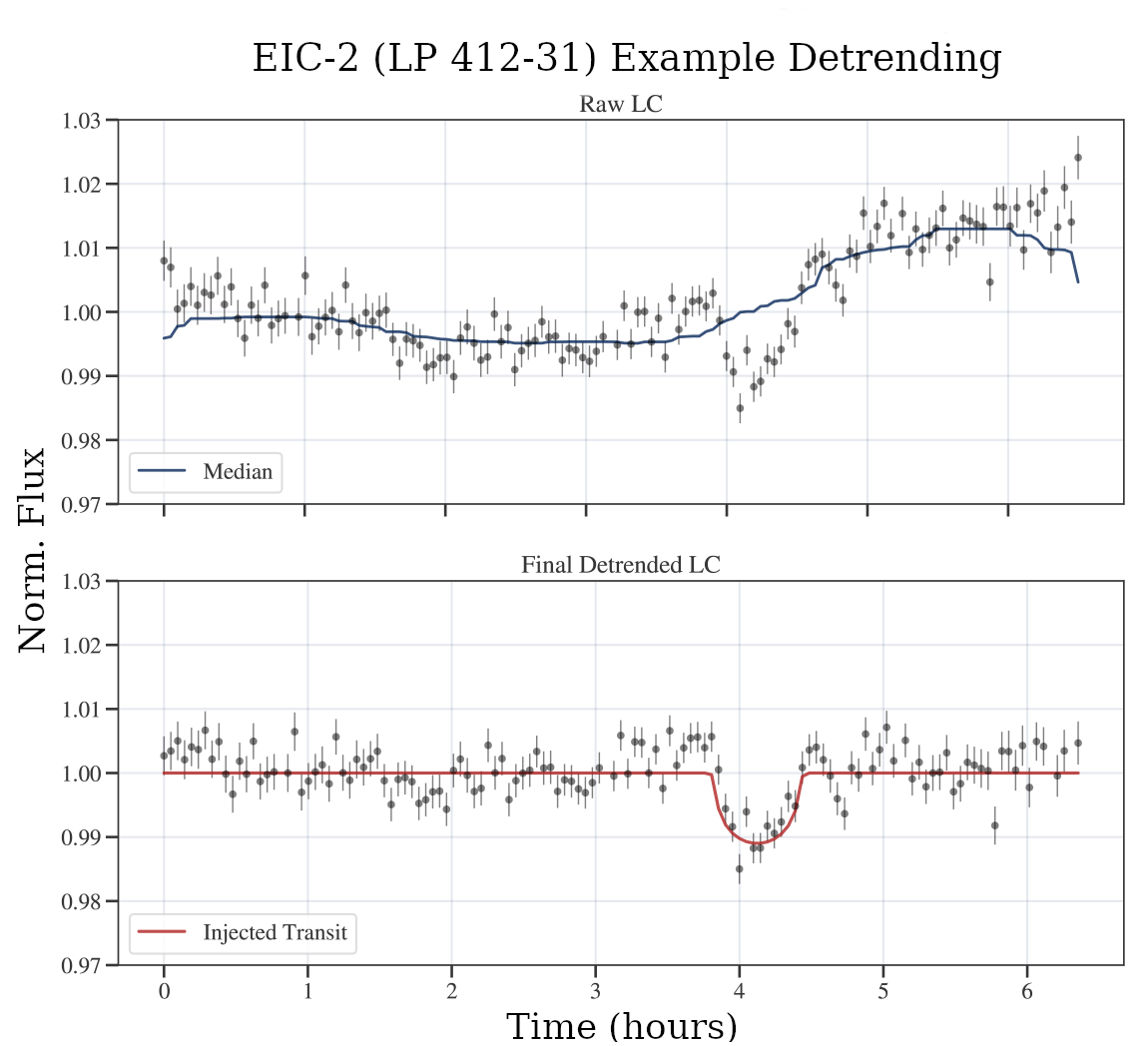}
\caption{\textbf{EIC-2 (LP 412-31) Example Detrending.} Data was taken with the Cassini telescope on 2018-12-11. The red line at bottom shows the injected transit signal (depth $\sim 1\%$, TRAPPIST-1~b orbit, with limb-darkening from \citet{Claret1998}) compared to the lightcurve after median detrending has been applied. The median shown at top is affected by some points outside the flux range.}
\label{fig:exdetrend}
\end{figure*}

After visual inspection, lightcurves undergo automated data cleaning and detrending. We fit a long-term trend with a median filter of two hours and $2\sigma$-clip upper outlying data before dividing out the trend. We do not clip below the median because of the risk of clipping deep transits. Median filtering will reduce the depth of all transits slightly, though our use of a two-hour filter window minimizes this effect for transits with durations of less than one hour, which comprises most of our discovery space. An example of median detrending applied to a real EDEN lightcurve with an injected transit of $\sim 1\%$ depth and TRAPPIST-1~b parameters is shown in Figure \ref{fig:exdetrend}.

While median detrending is a simple method, its effects are predictable and robust. Although the median filtering will not remove short-period, transit-like trends, it will not remove real transits either, if they are deeper than a few tenths of a percent (a danger of more complicated detrending techniques). Other trend-fitting methods with which we have experimented when performing transit injection tests include Savitzky-Golay \citep{Savitzky1964}, biweight, and multivariate polynomials constructed from external parameters such as airmass, and centroid positions. Savitzky-Golay and biweight filtering have very similar results to median detrending, and while multivariate polynomials can outperform median filters, they are also more likely to accidentally remove a real transit feature. Despite their relative simplicity, median filters are reliable \citep{Hippke2019wotan}.

\subsection{Transit Least Squares} \label{sec:TLS}

To search for transits in our detrended lightcurves, we utilize the package \texttt{Transit Least Squares} ({\TLS}, \citealt{Hippke2019}). The primary improvement over box least squares (BLS, \citealt{Kovacs2002}) is that rather than fitting a boxcar model to a time series, {\TLS} fits a more realistic, fixed transit shape with limb-darkening included, but the same parameters as BLS otherwise. We optimize the {\TLS} algorithm for our search by setting upper and lower limits on the stellar radius and mass to those for M dwarfs ($0.1 - 0.6 \mathrm{R_\odot}$, $0.08 - 0.5 \mathrm{M_\odot}$) and the maximum period to correspond to the approximate outer edge of the habitable zone (${\sim} 10$ days). We rely on our previously described data cleaning and detrending steps to remove bad data, and all lightcurves are weighed equally regardless of photometric precision. For each search we save a median-smoothed periodogram, as well as the phase folded model, transit parameters, false alarm probability (FAP), and signal detection efficiency (SDE) for the highest power period. 

\subsection{Candidate Vetting} \label{sec:vetting}

Most transit candidates identified by {\TLS} are false positives---and often obvious ones. Currently, vetting is done manually, but it may be automated in the future. The first check of a candidate is inspection of the viability of the {\TLS} output: are the transit parameters physical, does the phase folded lightcurve have obvious flares or systematic trends, what are the SDE and FAP values? If these are viable, the interactive data viewer is used to look at systematic trends during transit times, which usually reveal systematic noise sources that introduced the feature. We pursue follow-up observation to eliminate astrophysical false positives (such as eclipsing binaries) only after identifying a promising transit candidate not explainable by other means. We do not specifically set SDE or FAP values to eliminate transit candidates, and consider even those with poor statistics. However, we do perform an analysis of the SDE and FAP values that indicate a robust detection in Section \ref{sec:limits}.  

\section{The First EDEN Targets} \label{sec:targets}

 EIC-1 (\textit{\object{2MASSI J1835379+325954}}), EIC-2 (\textit{\object{LP 412-31}}), and EIC-3 (\textit{\object{2MUCD 20263}}) are all nearby M8/8.5 ultracool dwarfs (Table~\ref{tab:targetinfo}). 
 They are near the hydrogen burning limit and thus may be either high-mass brown dwarfs or low-mass stars. In this section we will briefly describe their stellar properties and past observations relevant to a search for planets. 

\subsection{EIC-1}

\textit{2MASSI J1835379+325954}, hereafter EIC-1, is an M8.5V dwarf located 5.7\,parsecs away \citep{Reid2003}. It was discovered and identified as a nearby dwarf by \citet{Lepine2002} as part of the Digitized Sky Survey. Its brown dwarf status is currently unknown due to differing lines of evidence \citep{Saur2018, Basri2009, Berdyugina2017}. It is a known radio pulsator with a strong magnetic field and a rapid $2.84$\,hr rotation period \citep{Berger2008,Berdyugina2017,Kuzmychov2017,Hallinan2008,Hallinan2015}. A possible detection of auroral emission has recently been reported for this target \citep{Hallinan2015}. 

EIC-1 has been the target of radial velocity (RV) observations by CARMENES \citep{TalOr2018} and Keck NIRSPEC \citep{Tanner2012}, some photometric monitoring by MEarth \citep{Dittman2016}, a wide-orbiting companion search by {\Spitzer} IRAC \citep{Carson2011}, and Subaru adaptive optics (AO) observations \citep{Siegler2005}, as well as numerous spectroscopic studies from UV to radio wavelengths. We are unaware of any companion candidates from these observations, but note that CARMENES identified it as ``active RV-loud'', potentially making the detection of habitable planets difficult by RV. EIC-1 was not observed by {\ktwo} and is scheduled to be observed by {\TESS} in Sector 26 in June 2020. 

\subsection{EIC-2}

\textit{LP 412-31}, hereafter EIC-2, is an M8V dwarf located 14.7\,parsecs away, identified by \citet{Kirkpatrick1995}. It has a rotational period of 0.61 days \citep{Irwin2011} and is a known flare star with a previously observed giant flare by XMM-Newton \citep{Stelzer2006}. 

EIC-2 has been the target of RV observations by the Red-Optical Planet Survey \citep{Barnes2014} and Keck NIRSPEC \citep{Rodler2012,Tanner2012}, which have $2\sigma$ sensitivity to $M\sin{i}>3.0M_\Earth$ throughout the habitable zone. It has also had periodic observations by MEarth \citep{Dittman2016}. It was not monitored by {\ktwo} or {\Spitzer} and is not scheduled to be observed by {\TESS} until after the primary mission due to its location near the ecliptic.

\subsection{EIC-3}

\textit{2MUCD 20263}, hereafter EIC-3, is an M8 dwarf located 15.6\,parsecs away, identified by \citet{Lepine2005}. Compared to EIC-1 and EIC-2, it has been the target of relatively few observations. It has been observed as part of MEarth and the SDSS-III APOGEE Radial Velocity Survey \citep{Deshpande2013}. It was not observed by {\ktwo} or {\Spitzer} and is scheduled to be observed by {\TESS} in Sector 20 in January 2020.

\begin{deluxetable*}{P{1cm}P{1.5cm}P{1cm}P{1cm}P{1cm}P{1cm}P{2cm}P{2cm}P{2cm}}[t]
\tablecaption{EDEN Targets \label{tab:targetinfo}}

\tablehead{\colhead{ID} & \colhead{Name} & \colhead{Spec. Type} & \colhead{Dist. (pc)} & \colhead{$I$ Mag} & \colhead{$K$ Mag} & \colhead{R.A. (J2000)} & \colhead{Decl. (J2000)} } 

\startdata
EIC-1 & 2MASSI J1835379 +325954 & M8.5V & 5.7 & 13.46 & 9.17 & 18:35:37.88 & +32:59:53.31 \\
\hline
EIC-2 & LP 412-31 & M8V & 14.7 & 14.48 & 10.64 & 03:20:59.71 & +18:54:22.77  \\
\hline
EIC-3 & 2MUCD 20263 & M8 & 15.6 & 14.35 & 10.84 & 07:14:03.94 & +37:02:46.03  \\
\hline
\enddata

\end{deluxetable*}

\section{Planet Detection Limits for EIC-1, EIC-2, and EIC-3} \label{sec:limits}

In this section we report the results of our previously described observations, data reduction and detrending pipelines, and transit search for the first three EDEN targets. Both visual and automatic transit injection and recovery tests are performed, described in Sections \ref{sec:visual_tests} and \ref{sec:injection_tests} respectively. We do not detect any convincing planet candidates for these stars, but place sensitive upper limits on the presence of transiting planets around them.

\subsection{Description of Lightcurves} \label{sec:lcs}

EIC-1, EIC-2, and EIC-3 were observed for 200 to 300 hours each from June 2018 to February 2019, with 40 to 60 individual observations per target (see Table~\ref{tab:logofobs}). The observations are highly clustered in time, with a few periods of continuous or nearly-continuous observations at different observatories lasting 24 hours or more. 

\begin{figure*}[t]
\centering
\includegraphics[width=1.0\textwidth]{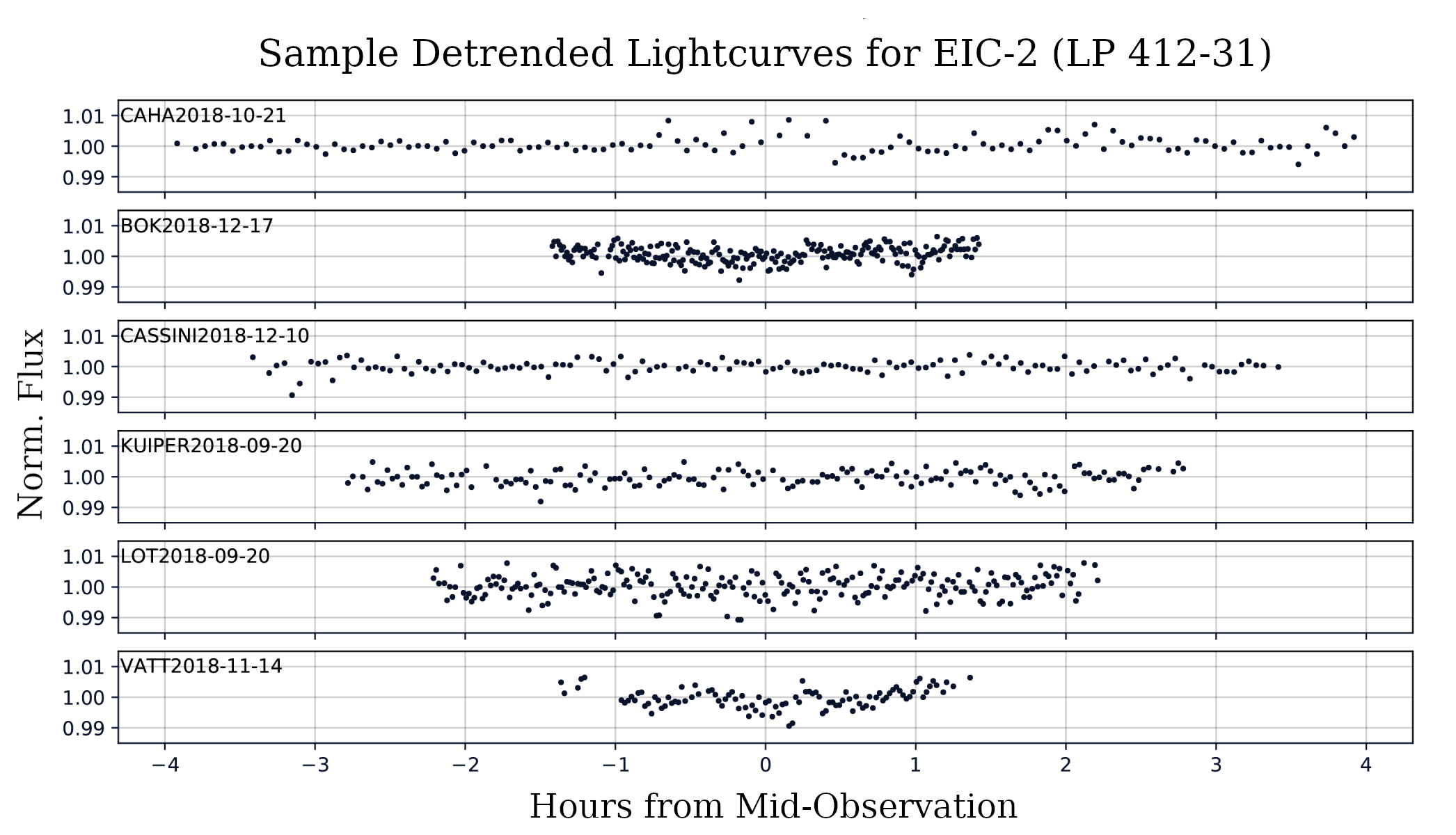}
\caption{EIC-2 (LP 412-31) Sample Lightcurves. The data are unbinned so that the relative cadence and raw precision of the instruments can be seen. Telescope and date are shown in the top left for each lightcurve.}
\label{fig:sample_lcs}
\end{figure*}

Roughly 60--80\% of the cleaned, detrended data are of sufficient quality for a subsequent transit search; the rest is affected by bad weather conditions or technical issues.
Durations for the individual high-quality lightcurves range between 2 and 10 hours, depending on target priority, observability, and weather. Some gaps less than 2 hours long exist within longer lightcurves because of passing clouds, temporary technical issues, or manual removal of flares or poor data sections. Cadences vary by a factor of ${\sim}$2--3 depending on the telescope (with higher cadence for larger primary mirrors) and detector readout times. The average median unbinned precision for lightcurves on a target is ${\sim} 0.28\%$. Trends are variable, but most lightcurves have nearly linear or parabolic variations of 1--3\% over their duration, possibly attributable to changing airmass or ponting drift. A sample of detrended lightcurves for EIC-2 for each telescope is shown in Figure \ref{fig:sample_lcs}.

Each target shows evidence for stellar activity, which is expected given their spectral type and previous observations described in Section \ref{sec:targets}. EIC-2 and EIC-3 have occasional flaring activity above $1\%$. Lightcurve segments with clearly-identifiable flares were removed manually before the transit search. Less than five flares were removed for both targets, representing a negligible loss in time. EIC-1 exhibits regular variability with a $0.5\%$ to $1\%$ amplitude, consistent with the rotational period of ${\sim} 3$\,hr \citep{Berger2008}. This variation can mimic transit-like signals, and thus reduces our transit detection sensitivity for the target.

\begin{deluxetable*}{P{1cm}P{1.5cm}P{1cm}P{1cm}P{3cm}P{1cm}}[t]
\tablecaption{Log of Observations \label{tab:logofobs}}

\tablehead{\colhead{ID} & \colhead{Name} & \colhead{Nights Obs.} & \colhead{Hours Obs.} & \colhead{Median Unbinned Precision ($\%$)} & \colhead{$\%$ used for {\TLS}}} 

\startdata
EIC-1 & 2MASSI J1835379 +325954 & 57 & 205.3 & 0.163 & ${\sim}70$ \\
\hline
EIC-2 & LP 412-31 & 56 & 311.7 & 0.315 & ${\sim}70$ \\
\hline
EIC-3 & 2MUCD 20263 & 43 &  297.5 & 0.380 & ${\sim}85$ \\
\enddata
\tablecomments{Appendix A provides a detailed log of the observations.}
\end{deluxetable*}

\subsection{Sensitivity Analysis} \label{sec:analysis}

To assess the transit detection capability of our observations, we implement a transit injection and recovery routine. We inject realistic transits into our raw target lightcurves using the analytic solutions of \citet{Mandel2002} as implemented in \texttt{batman} (BAsic Transit Model cAlculatioN, \citealt{Kreidberg2015}), re-select comparison stars with the same procedure described in Section~\ref{sec:compdetrend}, and attempt to recover the transit signals using our detrending and transit search pipeline. We also perform a limited visual transit recovery test to compare the sensitivity of the pipeline to a manual search by eye.

\subsubsection{Manual Transit Search} \label{sec:manual}

Before injecting any simulated transits, we perform a {\TLS} search and manual inspection of the lightcurves for each target to attempt to identify real transit candidates. Three team members reviewed every lightcurve individually and marked features of interest (transit candidates), which were then compared and vetted together according to Section \ref{sec:vetting}, along with the transit candidates identified by {\TLS}. We do not consider any of the transit candidates to be likely planets worthy of follow-up observation; we instead find them to be consistent with stellar variability and systematics. These steps do not definitively exclude the presence of transiting planets, but the probability of detecting a transiting planet is low, and will be quantified through our sensitivity analysis.

\subsubsection{Visual Transit Recovery Tests} \label{sec:visual_tests}

As a  comparison to the following {\TLS} sensitivity results in Section \ref{sec:injection_tests}, we also performed a limited, visual transit injection and recovery test. The purpose was to probe what transits team members could find by eye, without prior knowledge of their existence or location. 

One team member injected a TRAPPIST-1 b analog (1.1\,$R_{\Earth}$, $\sim 0.7\%$ depth, 1.51 day period, \citealt{Gillon2017}) at a random phase into a fraction of the lightcurves of each target (see Section  \ref{sec:injection_tests} for other parameters). Three other team members each received independent sets of these lightcurves with injections at random phase. Nearly half of the lightcurve sets did not contain any injections so that the team would not be compelled to identify transit candidates if they believed none were convincing. 

True positives are defined as real injections that are correctly identified, false positives are non-injection features wrongly identified as transits, and false negatives are real injections not identified. Collectively, out of 41 observed injected transits in 5 different sets of lightcurves, the team had a 1:1 true to false positive ratio, and a 4:1 false negative to true positive ratio. To determine our average visual sensitivity to TRAPPIST-1 b analogs, we consider how many sets of target lightcurves (containing multiple observed transit injections) had at least one true positive, irrespective of false negatives. Four out of five sets of target lightcurves with injections had one or more true positive, therefore we consider our average visual sensitivity to TRAPPIST-1 b analogs to be $\sim 80\%$. We believe this is limited by conservative transit identification rather than poor data quality. In reality, the false negative ratio is not as high as 4:1 since some of the ``observed'' transit injections are essentially unidentifiable due to only a small fraction of the transit being observed. While these tests are not a rigorous assessment of our ability to detect transits by eye, they support our argument that Earth-size planets can be correctly identified in our lightcurves without relying on the automated search routine.

\subsubsection{Automated Transit Recovery Tests} \label{sec:injection_tests}

The purpose of our automatic transit injection and recovery tests is to provide a scalable and objective sensitivity analysis method. For these tests, we simulate transits for planets in a logarithmic grid of period and radius from 0.5 to 10 days and 0.6 to 4 Earth radii, respectively, constituting most of our expected discovery space. The stellar radii of the target stars were determined from available surface gravity measurements \citep{Tsuji2016, Rajpurohit2018}. Within the grid, orbits are assumed to be circular with random phases and with impact parameters randomly drawn from a uniform distribution between $0.0$ and $1.0$. While it is technically feasible to detect transits up to an impact parameter of $1 + R_p$, $1.0$ is chosen as the upper limit since our detrending and search pipeline is not optimized to search for the very short duration and altered limb-darkening of grazing transits, and will have reduced sensitivity in that parameter space. We empirically find that sensitivity begins to drop significantly around impact parameters of 0.9, with around half the sensitivity to impact parameters between 0.9 and 1.0 compared to the average sensitivity below 0.9. Furthermore, a limit of $1 + R_p$ creates an artificial dependence on planet radius for transit sensitivity analysis, which distracts from more meaningful sensitivity trends.

The transit injections have quadratic limb darkening laws from \citet{Claret1998} for the $I$ band. While other limb darkening laws (e.g., logarithmic or exponential) may be more realistic \citep{Espinoza2016}, the differences for the sensitivity analysis are negligible in the present noise level regime. We further assume that the limb-darkening laws will be similar in all our NIR and red filters, and thus we use the same law for every injection.

Planets are injected at each grid point until there are 10 potentially detectable planets, i.e., planets with at least one simulated transit within the observing windows. 
This procedure is adopted to have a sufficient number of detectable planets at longer periods for counting statistics, where many planets may have no observed transits based on their random phase. Combining the grid size (12 by 8) with the requirement of 10 detectable planets means that, for each target, there are a total of 960 potentially recoverable transiting planets in the global sensitivity map.

\subsubsection{Positive Identification of Transits} \label{sec:transit_identification}

For us to consider a transit detected by {\TLS} to be a true positive result, it must meet one of the following two criteria:
(1) the best period is less than 0.5 hours different from the true period of the injected planet, or (2) at least one identified transit midpoint time is within 20 minutes of a real injected transit midpoint (i.e., a transit candidate was correctly identified, but the period is incorrect). All candidates which meet condition one, naturally meet condition two. 

We make an additional distinction between true positives recovered by {\TLS} and ``successful recoveries'', which we count in our sensitivity analysis. 
Successful recoveries are a subset of true positives that also pass a detection significance criterion.
We make this distinction because it is possible in a real search to detect a shallow transit only to dismiss it due to low signal.
We do not want to consider these cases as successful in our analysis.
Therefore, we limit successful recoveries in this analysis to detections that exceed a minimum signal detection efficiency (SDE) \citep{Hippke2019}, corresponding to a detection in a real search that would likely pass vetting and trigger follow-up observations. The SDE is the significance of a period relative to the average significance of all other periods.

We determine the minimum SDE for each target individually based on the global SDE distribution of false positives resulting from our injection recoveries. We set the minimum SDE required for a successful detection as the SDE that is greater than 95\% of false positives (i.e., only 5\% of false positives have a higher SDE). For our three targets, the minimum robust SDE value ranges for EIC-1, 2 and 3, are roughly 6, 7, and 11. 

\begin{figure*}[]
\centering
\includegraphics[width=1.0\textwidth]{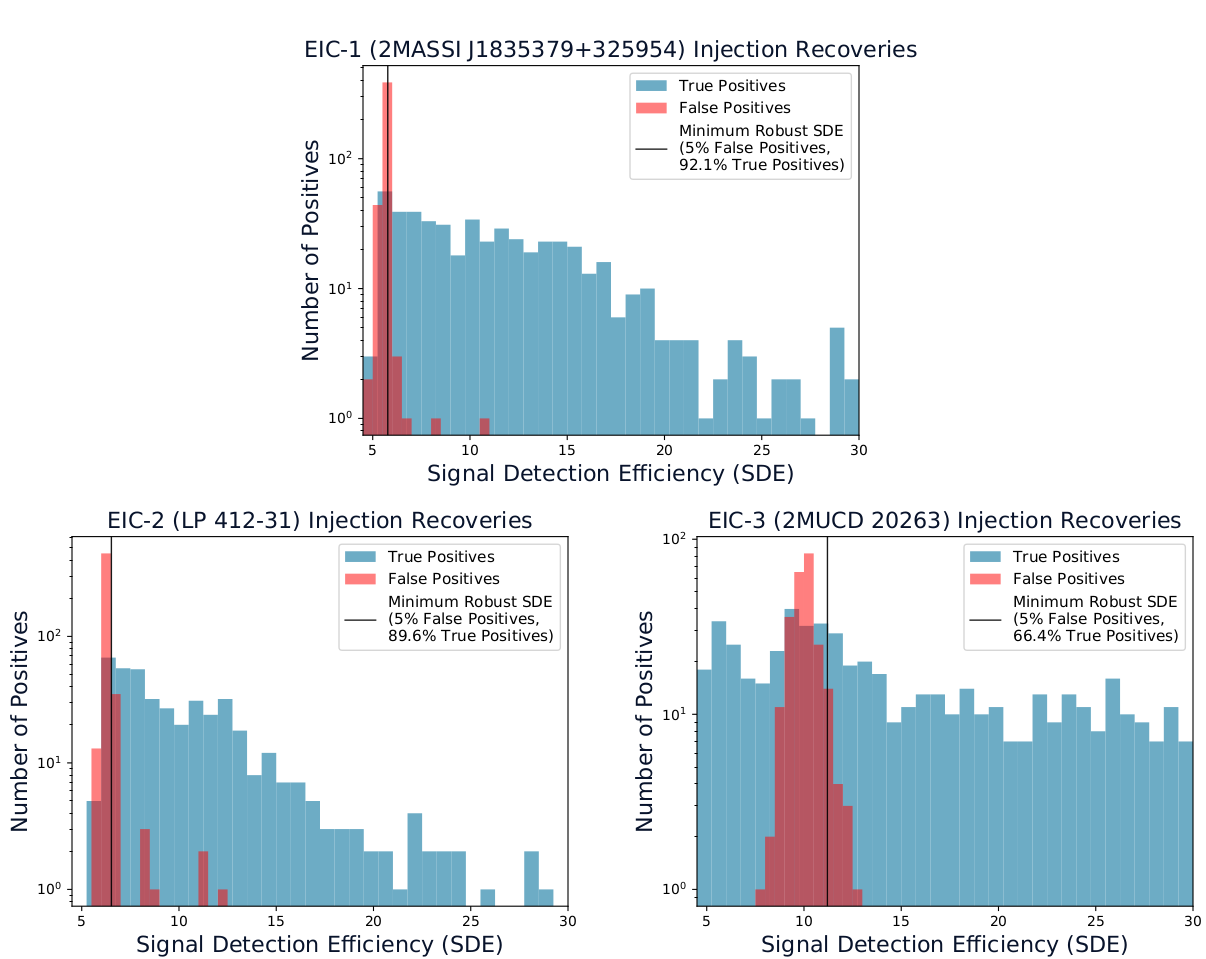}
\caption{\textbf{Signal Detection Efficiency (SDE) Distribution for EICs.} SDE is calculated as the signal to noise of the highest power in the recovery periodogram \citep{Hippke2019}. The number of false positives does not continue increasing for lower SDE values because of the characteristic false positive unique to each set of lightcurves. Further discussion can be found in Section \ref{sec:transit_identification}.}
\label{fig:T2_SDE}
\end{figure*}

The true and false positive distributions are shown for EIC-1, EIC-2, and EIC-3 in Figure \ref{fig:T2_SDE}. The differences result from the unique structure of each target's set of lightcurves, which produce higher and lower significance false positives. 
One noticeable feature of these plots (especially for EIC-3) is that the false positive distribution does not continually increase for lower SDE values, but is instead centered at a specific SDE. This potentially counter-intuitive distribution is caused by both the structure of each target's set of lightcurves, as well as the range and step size of the injection grid. Each target has a dominant false positive signal that is returned when there is no transit injection. Our grid range includes two rows of sub-Earth size planets that are extremely shallow in depth, and each injection in these rows will return nearly the same false positive SDE as if there was no injection. This leads to a build-up of a high fraction of false positives around the no injection SDE value, which corresponds roughly to the maximum of the false positive distribution. The higher fraction of true positives at lower SDE values is due to the fact that there is a certain range of injection depths that will only be a marginally higher power than the no injection false positive and thus will have a low SDE, but they will still be detected successfully at high rates. 

It is important to note that the SDE cutoff is not used to determine the significance of transit candidates in the real transit search and is only used in finding the significance of injection recoveries after concluding by other means (Section \ref{sec:manual}) that the data contains no real transits. Therefore, it likely provides a conservative sensitivity estimate. Finally, the SDE cutoff cannot be expected to fully capture the probability that a true positive candidate would be followed-up and confirmed, but rather is a best attempt at conservatively estimating the likelihood given subjective human involvement in deciding what is and what is not a convincing candidate. While it would be more desirable to build a completely automatic vetting algorithm, for our observations the algorithm would need to be prohibitively intelligent and complex, and could result in more missed planets.

\subsubsection{Pipeline Sensitivity} \label{sec:pipeline_sensitivity}

We illustrate our transit detection sensitivity for EIC-1, EIC-2, and EIC-3 
in
Figures \ref{fig:2M1835}, 
\ref{fig:LP412-31}, and 
\ref{fig:2MUCD20263},
respectively. The top plots show the efficiency of our pipeline to detect transiting planets, while the bottom plots represent total detection probability for all planets, both transiting and non-transiting, based on our transit detection sensitivity and the geometric transit probability for planets as a function of semi-major axis ($P_{tr} = \frac{R_*}{a}$). To calculate the overall sensitivity within a specific range of periods and radii, we simply average the detection sensitivity in that range. Mean sensitivities for select ranges are shown in Table~\ref{tab:sensitivity}.

\begin{deluxetable*}{@{\extracolsep{4pt}}ccccccc}[t]
\tablecaption{EDEN Sensitivity \label{tab:sensitivity}}

\tablehead
{
\colhead{}&
  \multicolumn{3}{c}{Transit Sensitivity (\%)}&
  \multicolumn{3}{c}{Total Detectability (\%)} \\
\cline{2-4} \cline{5-7} 
\colhead{ID}& \colhead{0.5 to 2 days}& 
\colhead{2 to 5 days}& \colhead{5+ days}& 
\colhead{0.5 to 2 days}& 
\colhead{2 to 5 days}& \colhead{5+ days}
}
\startdata
EIC-1 & 60 $\pm$ 10 & 35 $\pm$ 5 & 12 $\pm$ 1 & 4.0 $\pm$ 0.5 & 1.1 $\pm$ 0.25 & 0.18 $\pm$ 0.05 \\
EIC-2 & 40 $\pm$ 10 & 22 $\pm$ 5 & 10 $\pm$ 1 & 2.5 $\pm$ 0.5 & 0.6 $\pm$ 0.25 & 0.17 $\pm$ 0.05\\
EIC-3 & 80 $\pm$ 10 & 40 $\pm$ 5 & 8 $\pm$ 1 & 5.3 $\pm$ 0.5 & 1.1 $\pm$ 0.25 & 0.13 $\pm$ 0.05 \\
\enddata
\tablecomments{Reported transit sensitivity and total detectability values are averages for planets between one and two Earth radii. Listed errors are the standard error of the mean.}
\end{deluxetable*}

\newpage
\section{Discussion}\label{sec:discussion}

\subsection{EDEN Sensitivity}\label{sec:sensitivity}

The sensitivity maps for EIC-1, EIC-2, and EIC-3 show that we have the potential to successfully detect transiting Earth-sized planets in the habitable zones of nearby, ultracool dwarfs. Furthermore, they show that in a few cases we can detect sub-Earth-sized planets on closer orbits provided two or more transits occur during high-quality observations. To compare these results with {\TESS}, the estimated photometric precisions for EIC-1, EIC-2, and EIC-3 are 0.136, 0.299, and 0.343 $\%$ respectively in \em{one hour}\em{}
periods of observation ({\TESS} Mag. 13.28, 14.35, and 14.52) \citep{Stassun2018}. These are very similar to the median achieved precisions of unbinned EDEN lightcurves typically at a \em{one minute}\em{} cadence (0.163, 0.315, and 0.380 $\%$ respectively). Thus, with long-term targeted observations it is possible we could achieve better sensitivities than {\TESS} for single targets, in cases where the benefit of our increased photometric precision can outweigh the benefit of {\TESS}'s continuous 28 day coverage. 

\subsection{Sensitivity Analysis and Detection Biases} \label{sec:analysis_bias}

The primary goal of our sensitivity analysis is setting planetary limits around the target stars that will be useful for future observations. These limits can potentially improve the efficiency of similar transit surveys, and in the case of any future radial velocity (RV) companion candidates, help to constrain the inclination. The secondary goal is to help to identify strengths, weaknesses, and biases of our observations and routines. Using this information we can improve our future observations, data reduction, detrending, and search methods. That being stated, we believe our methods are nearly optimized, and only minor improvements can still be expected, which would not significantly change our sensitivity results.

\begin{figure*}[p]
\centering
\includegraphics[width=0.85\textwidth]{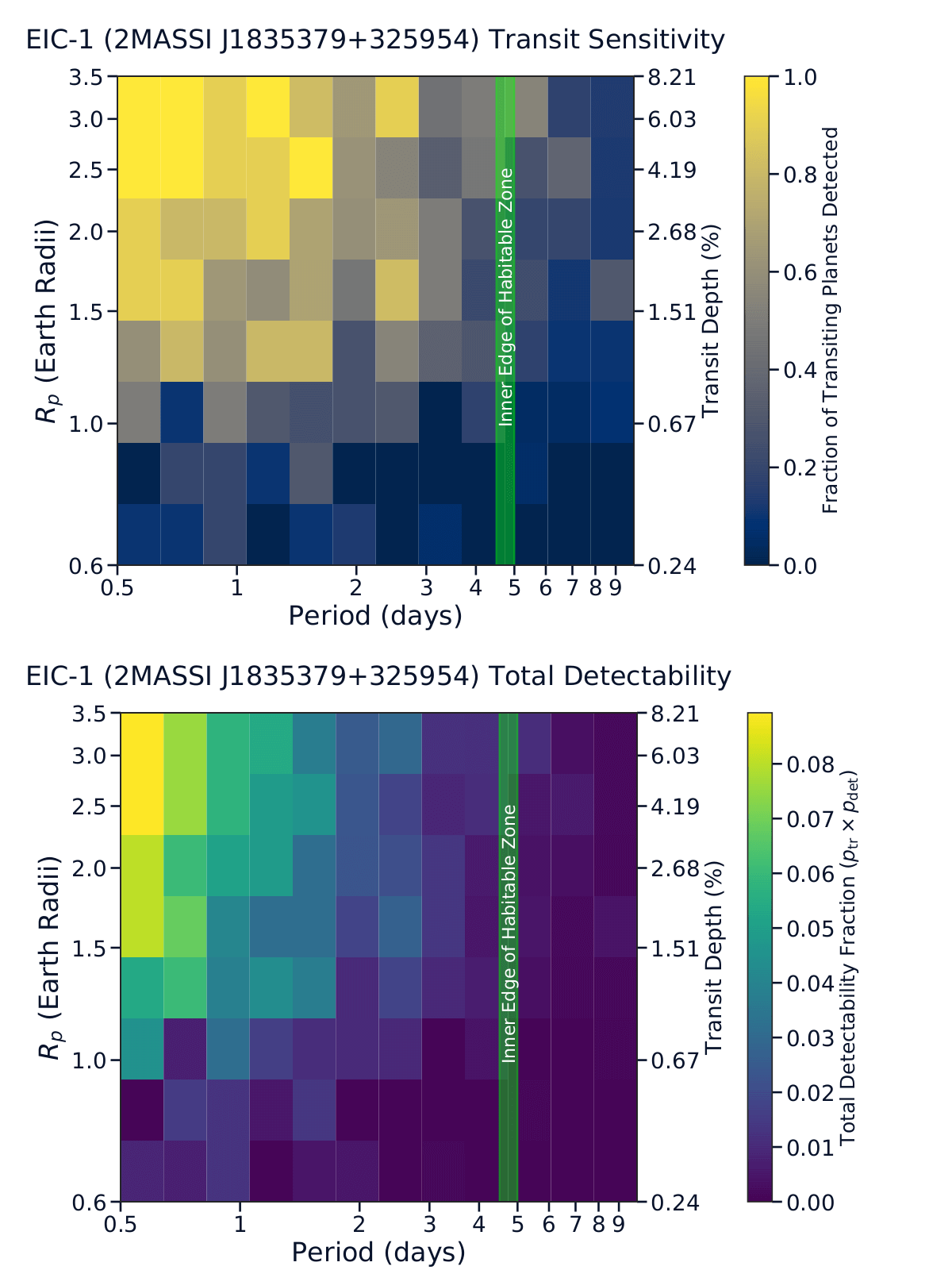}
\caption{\textbf{EIC-1 (2MASSI J1835379+325954) sensitivity maps.} Top: Pipeline sensitivity to transiting planets. Each grid block represents the fraction of transiting planets recovered out of all injected planets (both recoverable and non-recoverable) for a period and radius centered within the block. Bottom: Total detectability considering the geometric transit probability ($p_{tr} \times p_{det}$).}
\label{fig:2M1835}
\end{figure*}

\begin{figure*}[p]
\centering
\includegraphics[width=0.85\textwidth]{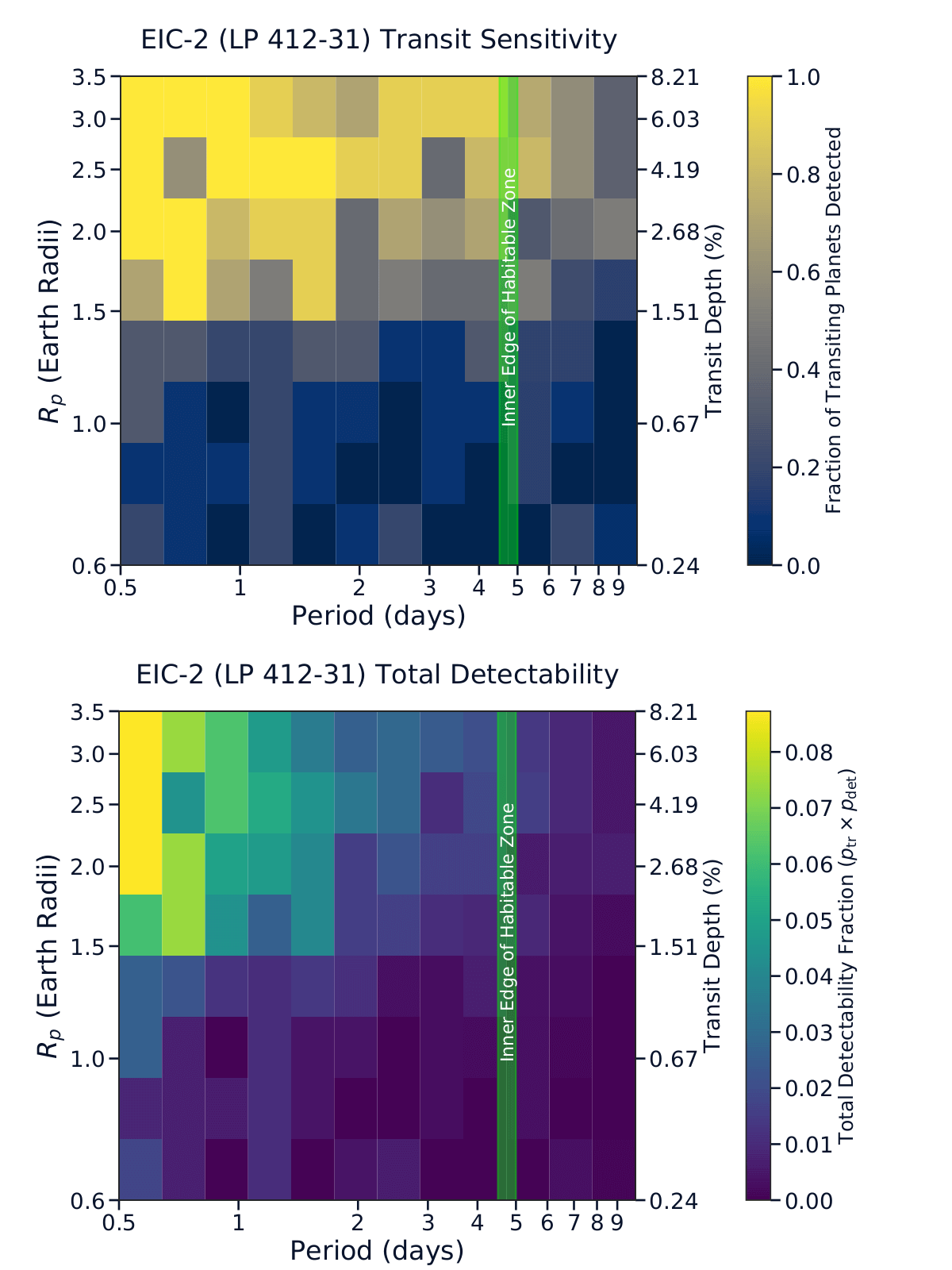}
\caption{\textbf{EIC-2 (LP 412-31) sensitivity maps.} Top: Pipeline sensitivity to transiting planets. Each grid block represents the fraction of transiting planets recovered out of all injected planets (both recoverable and non-recoverable) for a period and radius centered within the block. Bottom: Total detectability considering the geometric transit probability ($p_{tr} \times p_{det}$).}
\label{fig:LP412-31}
\end{figure*}

\begin{figure*}[p]
\centering
\includegraphics[width=0.85\textwidth]{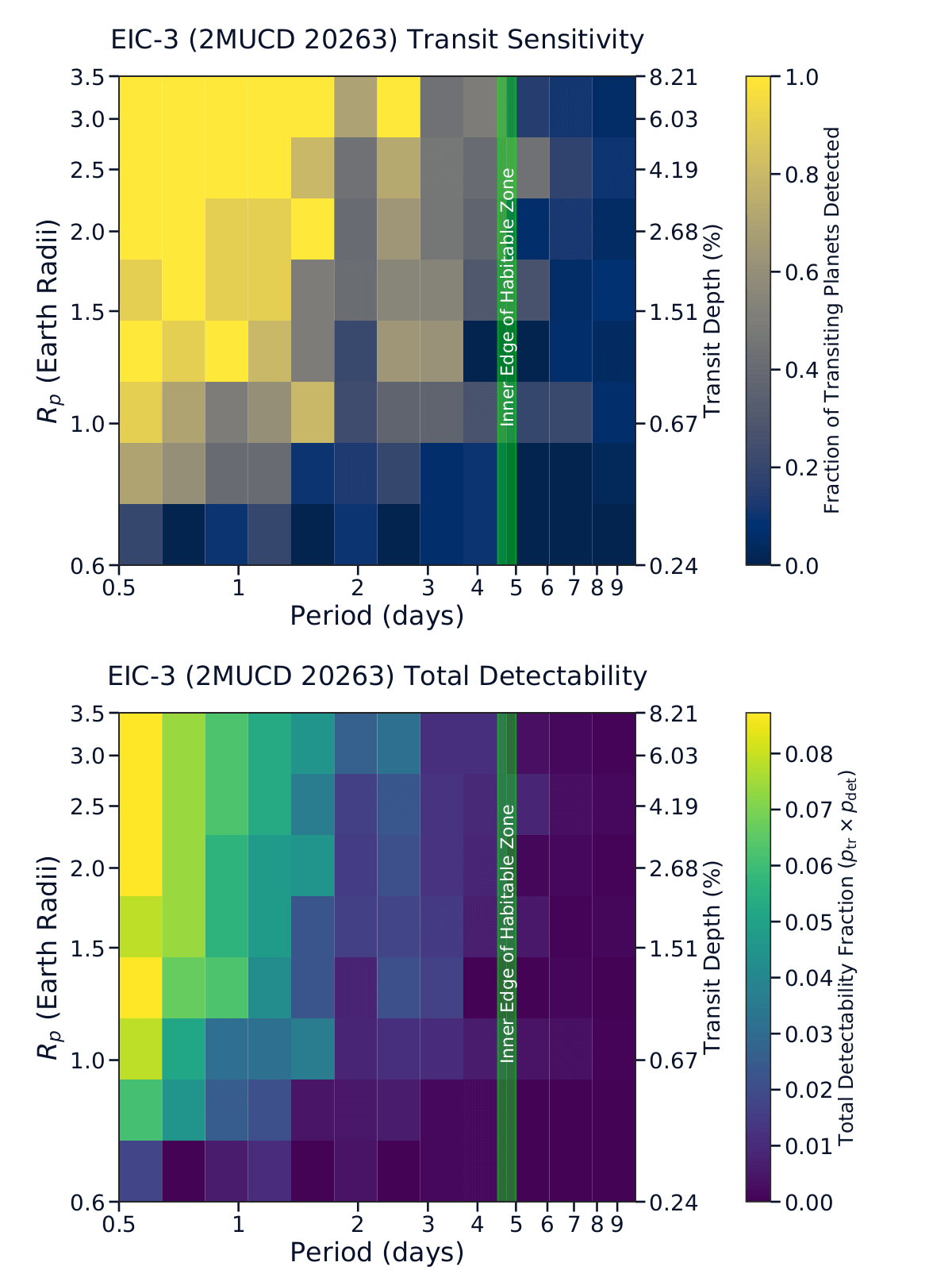}
\caption{\textbf{EIC-3 (2MUCD 20263) sensitivity maps.} Top: Pipeline sensitivity to transiting planets. Each grid block represents the fraction of transiting planets recovered out of all injected planets (both recoverable and non-recoverable) for a period and radius centered within the block. Bottom: Total detectability considering the geometric transit probability ($p_{tr} \times p_{det}$).}
\label{fig:2MUCD20263}
\end{figure*}

The sensitivity maps in Figures \ref{fig:2M1835}, \ref{fig:LP412-31} and \ref{fig:2MUCD20263}
show two distinct gradients of decreasing sensitivity. As one would expect, these gradients are for smaller planets ($<1 R_\Earth$, i.e., lower transit signal-to-noise), and longer periods ($>3$ days, i.e. fewer observed transits). Both regions of low sensitivity have more true positives than are considered successful, since many detections will have low significance that may not be followed-up. It is possible that some of these true positives would be followed-up, therefore it is likely that the map is somewhat conservative. Furthermore, our manual injection and recovery by eye test estimated that our sensitivity to TRAPPIST-1~b analogs is $\sim 80\%$, while the average automated sensitivity is $\sim 30\%$. This provides additional evidence that the automated sensitivity is conservative, especially for longer period planets where one transit can be successfully detected by eye. As a final point, on the right side of the bottom plot of Figures \ref{fig:2M1835}, \ref{fig:LP412-31} and \ref{fig:2MUCD20263}, where geometric probability is considered, the gradient for longer period planets becomes steeper, reflecting the decreasing transit probability at greater distance from the host star.

One noticeable aspect of our sensitivity maps is higher noise than similar plots from space-based missions. The noise is due to four primary factors, including the limited grid size, random transit times, the relatively low number of planets injected, as well as the sporadic and discontinuous schedule of EDEN observations. Most single blocks with relatively high or low sensitivity are simply due to the random sample times. Unlike observations from \textit{Kepler}, it is possible that by misfortune a short-period planet never transits during an observation. Some columns may also have lower or higher sensitivity compared to their surroundings depending on whether or not the period is close to a harmonic of the period of observations, and therefore are more or less sensitive to phase. 

\subsection{Inner Planets and Outer Planets}

Our detection limits for inner, shorter period planets can place significant constraints on the probability of outer, longer period planets, where observational coverage is lacking, in light of the occurrence rates of small planets around M-dwarfs \citep{Mulders2015a}. The strongest example of this is the TRAPPIST-1 system. TRAPPIST-1b and c were detected by ground based observations that motivated space-based follow-up, which discovered longer-period planets. For our targets, the approximate probability to detect transiting planets analogous to TRAPPIST-1b and c with \textit{one or more transits} is $\sim50\%$. The lack of close-in transiting planets in the extensive datasets on our targets decreases the probability that there are transiting planets at longer periods, and suggests continued observation to increase sensitivity for them is not be pragmatic, given the much larger volume of data needed.

\subsection{Constraints on Planet Formation Theory}
The sample of planets around very cool stars is still small, since late M-dwarfs are too faint for wide-field transit surveys. In addition, higher stellar activity can further complicate the analyses of their lightcurves \citep[e.g.,][]{Perger2017}. EDEN has unique capabilities to target these stars and any planet our survey may detect will serve as a valuable addition to this small sample. The examples of TRAPPIST-1 \citep{Gillon2017} and GJ~3512b \citep{Morales2019} showed how individual discoveries can challenge our current understanding of planet formation and inform tests of competing formation theories. To assess such discoveries in terms of the actual underlying population of exoplanets, it is crucial to be aware of and able to quantify the relevant selection biases. With a well-defined target selection function, an automated detection pipeline, and the thorough sensitivity analysis presented here, we are prepared to accurately model the EDEN selection biases. Correcting for these biases enables detailed occurrence rate measurements and builds the foundation to study the demographics of late M-dwarf planetary systems.

The inferred bias can also be applied to synthetic planets from a theoretical formation model. The resulting \textit{observable} synthetic population enables statistical comparisons between theory and observations \citep[e.g.,][]{Mordasini2009}. Detailed forward models of well-characterized exoplanet surveys can directly test planet formation models and even optimize free parameters \citep[][]{Mulders2018,Mulders2019}. Such dedicated M-dwarf population syntheses are  powerful tools to constrain planet formation in a parameter space different from that around solar-type stars. The predictive power of exoplanet surveys depends on the survey's sensitivity and the number of targets observed: as the number of targets observed by EDEN increases, the emerging planet statistics will increase in significance. Currently, we are surveying targets at an increasing rate.

\section{Conclusions} \label{sec:conc}

We present the first lightcurves and sensitivity analysis from the EDEN transiting exoplanet survey. The key results of our studies are as follows:

1) EDEN's 0.6--2.3\,m diameter telescopes provide very high-quality (median 0.28$\%$ precision) red-visual (500--900\,nm) lightcurves for late-M-dwarf stars in the solar neighborhood.

2) We present data on three nearby late-M dwarfs, obtained in the context of a multi-continental transit search campaign. Our observations include 57, 56, and 43 nights of data on the three targets (EIC-1, EIC-2, EIC-3), respectively.

3) We reviewed the EDEN data reduction and photometry pipeline and our de-trending and transit search procedure. Our procedure has been tested, optimized, and validated through transit injection-and-recovery tests.

4) Our lightcurves reach the sensitivity to detect transits of Earth-sized planets. In the total of 156 observations on the three targets, no convincing candidate transit events have been identified. 

5) We describe our transit injection-and-recovery-based approach to assess sensitivity to planetary transits as a function of planet radius and orbital period. We provide a detailed assessment of the sensitivity to transits around our three targets. We show these estimates are conservative compared to manual transit searches by eye.

6) Our data can confidently exclude the presence of Earth-sized transiting planets with orbital periods shorter than 1 day around each of the targets. Earth-sized planets with 1--2\,day periods would have been detected in our data in two transits with $\sim 60\%$ probability.

7) EDEN reaches a sensitivity to Earth-sized planets around faint red dwarf stars ($I\approx 15$ mag), which are challenging targets even for NASA's {\TESS} mission. Thus, EDEN data on such systems can provide complementary information to {\TESS} lightcurves.

8) Our study demonstrates the potential of the EDEN survey to robustly probe the presence of transiting, Earth-sized planets within and inside of the habitable zones of nearby late red dwarfs and, in case of non-detection, to set stringent upper limits on the presence of such planets.

\acknowledgments

The results reported herein benefited from collaborations and/or information exchange within NASA’s Nexus for Exoplanet System Science (NExSS) research coordination network sponsored by NASA’s Science Mission Directorate.
This study used results from the RECONS project (recons.org).
T. Henning acknowledges support from the European Research Council under the Horizon 2020 Framework Program via the ERC Advanced Grant Origins 83 24 28.
B.V. Rackham acknowledges support from the Heising-Simons Foundation. 

This research made use of Photutils, an Astropy package for
detection and photometry of astronomical sources \citep{Bradley2019}.

The principal investigators of EDEN are D. Apai, P. Gabor, T. Henning, and W-P. Chen. Initial target selection was performed by A. Mousseau, A. Bixel, and D. Apai. Telescope allocation is organized by D. Apai, L. Mancini, W-P. Chen, C.C. Ngeow, and P. Gabor.
Observations have been performed by H. Baehr, A. Bhandare. A. Bixel, R. Boyle (pilot studies as VATT's Telescope Scientist), S. Brown, J. Dietrich, A. Gibbs, M. H\"aberle, W. Ip, M. Keppler, L. Mancini, V. Marian, K. Molaverdikhani, A. Mousseau, J. Perez Chavez, B. Rackham, P. Sarkis, M. Schlecker, Q.J. Socia, A. Tsai and others. 
Software has been developed by A. Bixel, N. Espinoza, A. Gibbs, J. Perez Chavez, and B. Rackham. The EDEN automatic pipeline ({\edenAP}) was developed by N. Espinoza (precursor pipeline), B. Rackham (bulk development), A. Bixel, J. Perez Chavez (calibration steps), and A. Gibbs. Data organization and collection, and the interactive data viewer were developed by A. Bixel. Detrending, transit search, sensitivity analysis steps have been implemented and developed by A. Gibbs.

This research has made use of the Cassini 1.52\,m telescope, which is
operated by INAF-OAS ``Osservatorio di Astrofisica e Scienza dello
Spazio'' of Bologna in Loiano (Italy).

We thank the mountain operations staff at the University of Arizona, Mount Lemmon Sky Center, Lulin Observatory, Calar Alto Observatory, Loaino Telescopes, Mount Graham International Observatory, Vatican Advanced Technology Telescope, and Kitt Peak National Observatory.


\facilities{Mount Lemmon Sky Center, Lulin Observatory, Calar Alto Observatory, Loiano 152cm Cassini Telescope, Kuiper 61-inch Telescope, Vatican Advanced Technology Telescope (VATT), Bok 2.3m Telescope }


\software{Numpy \citep{vanderWalt2011}, Pandas\citep {Mckinney2010}, Scipy \citep{Virtanen2019}, Astropy \citep{astropy:2018}, Photutils \citep{Bradley2019}, astronomy.net \citep{Lang2010}, {\TLS} \citep{Hippke2019}, \texttt{batman} \citep{Kreidberg2015}, \texttt{edenAP}}



\appendix

\section{Observation Log}

In case of future research or discoveries where EDEN data may be useful, we list all periods of observations for EIC-1, EIC-2, and EIC-3 in Tables \ref{tab:t1obslist}, \ref{tab:t2obslist}, and \ref{tab:t3obslist}.


\bibliography{references}
\bibliographystyle{aasjournal}



\begin{deluxetable*}{P{2cm}P{2cm}P{2cm}P{2cm}P{1cm}}[t]
\tabletypesize{\scriptsize}
\tablecaption{EIC-1 (2MASSI J1835379+325954) List of Observations} \label{tab:t1obslist}
\tablehead{\colhead{Telescope} & \colhead{Local Date} & \colhead{BJD Start ($-245700$)} & \colhead{BJD End ($-245700$)} & \colhead{Hours} } 

\startdata
CAHA     & 2018-06-26 & 1296.354 & 1296.681 & 7.8 \\
CASSINI  & 2018-06-29 & 1299.354 & 1299.612 & 6.2 \\
CASSINI  & 2018-06-30 & 1300.342 & 1300.607 & 6.4 \\
CAHA     & 2018-06-30 & 1300.356 & 1300.433 & 1.9 \\
CAHA     & 2018-07-01 & 1301.357 & 1301.409 & 1.2 \\
CAHA     & 2018-07-02 & 1302.349 & 1302.407 & 1.4 \\
CASSINI  & 2018-07-02 & 1302.353 & 1302.379 & 0.6 \\
CAHA     & 2018-07-03 & 1303.354 & 1303.379 & 0.6 \\
CAHA     & 2018-07-04 & 1304.367 & 1304.652 & 6.8 \\
KUIPER   & 2018-07-18 & 1318.743 & 1318.959 & 5.2 \\
CASSINI  & 2018-07-19 & 1319.328 & 1319.615 & 6.9 \\
CAHA     & 2018-07-19 & 1319.346 & 1319.564 & 5.2 \\
KUIPER   & 2018-07-19 & 1319.868 & 1319.982 & 2.7 \\
CASSINI  & 2018-07-20 & 1320.328 & 1320.599 & 6.5 \\
CAHA     & 2018-07-20 & 1320.348 & 1320.658 & 7.5 \\
CASSINI  & 2018-07-21 & 1321.33  & 1321.562 & 5.6 \\
CAHA     & 2018-07-22 & 1322.351 & 1322.675 & 7.8 \\
CASSINI  & 2018-07-23 & 1323.332 & 1323.615 & 6.8 \\
CAHA     & 2018-07-23 & 1323.347 & 1323.454 & 2.6 \\
CAHA     & 2018-07-24 & 1324.346 & 1324.671 & 7.8 \\
CASSINI  & 2018-07-25 & 1325.5   & 1325.62  & 2.9 \\
CAHA     & 2018-07-25 & 1325.507 & 1325.671 & 3.9 \\
CAHA     & 2018-07-26 & 1326.508 & 1326.671 & 3.9 \\
LOT      & 2018-07-29 & 1329.147 & 1329.211 & 1.5 \\
KUIPER   & 2018-09-03 & 1365.658 & 1365.812 & 3.7 \\
KUIPER   & 2018-09-04 & 1366.687 & 1366.773 & 2.1 \\
KUIPER   & 2018-09-05 & 1367.711 & 1367.802 & 2.2 \\
KUIPER   & 2018-09-06 & 1368.624 & 1368.808 & 4.4 \\
KUIPER   & 2018-09-07 & 1369.6   & 1369.804 & 4.9 \\
CAHA     & 2018-09-16 & 1378.308 & 1378.493 & 4.5 \\
CAHA     & 2018-09-17 & 1379.299 & 1379.303 & 0.1 \\
BOK      & 2018-09-17 & 1379.643 & 1379.753 & 2.6 \\
LOT      & 2018-09-18 & 1379.995 & 1380.147 & 3.6 \\
LOT      & 2018-09-19 & 1381.022 & 1381.144 & 2.9 \\
LOT      & 2018-09-20 & 1382.006 & 1382.143 & 3.3 \\
CAHA     & 2018-09-20 & 1382.297 & 1382.472 & 4.2 \\
KUIPER   & 2018-09-20 & 1382.605 & 1382.777 & 4.1 \\
CAHA     & 2018-09-21 & 1383.288 & 1383.479 & 4.6 \\
CAHA     & 2018-09-22 & 1384.287 & 1384.486 & 4.8 \\
CAHA     & 2018-09-23 & 1385.304 & 1385.482 & 4.3 \\
CAHA     & 2018-09-24 & 1386.294 & 1386.463 & 4.0   \\
CAHA     & 2018-09-25 & 1387.299 & 1387.423 & 3.0   \\
SCHULMAN & 2018-09-25 & 1387.605 & 1387.714 & 2.6 \\
CAHA     & 2018-09-28 & 1390.321 & 1390.46  & 3.3 \\
SCHULMAN & 2018-09-28 & 1390.59  & 1390.729 & 3.3 \\
CAHA     & 2018-09-30 & 1392.278 & 1392.417 & 3.3 \\
CAHA     & 2018-10-17 & 1409.282 & 1409.405 & 3.0   \\
LOT      & 2018-10-19 & 1411.061 & 1411.086 & 0.6 \\
CAHA     & 2018-10-19 & 1411.341 & 1411.414 & 1.8 \\
LOT      & 2018-10-20 & 1411.964 & 1412.037 & 1.8 \\
CAHA     & 2018-10-22 & 1414.284 & 1414.364 & 1.9 \\
CAHA     & 2018-10-24 & 1416.265 & 1416.366 & 2.4 \\
CAHA     & 2018-10-25 & 1417.253 & 1417.37  & 2.8 \\
KUIPER   & 2018-10-31 & 1423.597 & 1423.644 & 1.1 \\
KUIPER   & 2018-11-03 & 1426.555 & 1426.645 & 2.2 \\
KUIPER   & 2018-11-04 & 1427.55  & 1427.642 & 2.2
\enddata

\end{deluxetable*}

\begin{deluxetable*}{P{2cm}P{2cm}P{2cm}P{2cm}P{1cm}}[t]
\tablecaption{EIC-2 (LP 412-31) List of Observations} \label{tab:t2obslist}

\tablehead{\colhead{Telescope} & \colhead{Local Date} & \colhead{BJD Start ($-245700$)} & \colhead{BJD End ($-245700$)} & \colhead{Hours} } 

\startdata
KUIPER  & 2018-09-04 & 1366.798 & 1367.012 & 5.1 \\
KUIPER  & 2018-09-05 & 1367.817 & 1368.024 & 5.0   \\
KUIPER  & 2018-09-07 & 1369.812 & 1370.013 & 4.8 \\
LOT     & 2018-09-17 & 1379.16  & 1379.372 & 5.1 \\
CAHA    & 2018-09-17 & 1379.48  & 1379.722 & 5.8 \\
BOK     & 2018-09-17 & 1379.776 & 1379.874 & 2.3 \\
LOT     & 2018-09-18 & 1380.156 & 1380.342 & 4.4 \\
CAHA    & 2018-09-18 & 1380.54  & 1380.719 & 4.3 \\
LOT     & 2018-09-19 & 1381.151 & 1381.338 & 4.5 \\
LOT     & 2018-09-20 & 1382.15  & 1382.339 & 4.5 \\
KUIPER  & 2018-09-20 & 1382.787 & 1383.028 & 5.8 \\
CAHA    & 2018-09-21 & 1383.505 & 1383.714 & 5.0   \\
CAHA    & 2018-09-22 & 1384.511 & 1384.716 & 4.9 \\
CAHA    & 2018-09-23 & 1385.502 & 1385.714 & 5.1 \\
CAHA    & 2018-09-24 & 1386.482 & 1386.616 & 3.2 \\
CAHA    & 2018-09-27 & 1389.545 & 1389.712 & 4.0   \\
CAHA    & 2018-09-28 & 1390.484 & 1390.718 & 5.6 \\
CAHA    & 2018-09-30 & 1392.433 & 1392.714 & 6.7 \\
LOT     & 2018-10-16 & 1408.199 & 1408.223 & 0.6 \\
CAHA    & 2018-10-17 & 1409.415 & 1409.636 & 5.3 \\
CAHA    & 2018-10-18 & 1410.632 & 1410.651 & 0.5 \\
LOT     & 2018-10-19 & 1411.096 & 1411.381 & 6.8 \\
CAHA    & 2018-10-19 & 1411.433 & 1411.505 & 1.7 \\
LOT     & 2018-10-20 & 1412.142 & 1412.391 & 6.0   \\
CAHA    & 2018-10-21 & 1413.392 & 1413.723 & 7.9 \\
CAHA    & 2018-10-22 & 1414.378 & 1414.58  & 4.9 \\
CAHA    & 2018-10-23 & 1415.392 & 1415.396 & 0.1 \\
CAHA    & 2018-10-24 & 1416.377 & 1416.734 & 8.6 \\
CAHA    & 2018-10-25 & 1417.38  & 1417.571 & 4.6 \\
KUIPER  & 2018-10-31 & 1423.662 & 1424.016 & 8.5 \\
KUIPER  & 2018-11-01 & 1424.66  & 1425.032 & 8.9 \\
KUIPER  & 2018-11-02 & 1425.687 & 1425.994 & 7.4 \\
KUIPER  & 2018-11-03 & 1426.658 & 1427.037 & 9.1 \\
KUIPER  & 2018-11-04 & 1427.683 & 1428.045 & 8.7 \\
KUIPER  & 2018-11-09 & 1432.627 & 1432.961 & 8.0   \\
KUIPER  & 2018-11-10 & 1433.64  & 1434.041 & 9.6 \\
KUIPER  & 2018-11-11 & 1434.605 & 1435.018 & 9.9 \\
VATT    & 2018-11-11 & 1434.632 & 1435.013 & 9.1 \\
VATT    & 2018-11-12 & 1435.665 & 1436.011 & 8.3 \\
VATT    & 2018-11-13 & 1436.664 & 1437.006 & 8.2 \\
VATT    & 2018-11-14 & 1437.84  & 1438.002 & 3.9 \\
VATT    & 2018-11-15 & 1438.656 & 1438.997 & 8.2 \\
VATT    & 2018-11-16 & 1439.767 & 1440.005 & 5.7 \\
VATT    & 2018-11-17 & 1440.709 & 1441.006 & 7.1 \\
VATT    & 2018-11-18 & 1441.665 & 1442     & 8.0   \\
VATT    & 2018-11-19 & 1442.654 & 1442.997 & 8.2 \\
CASSINI & 2018-11-28 & 1451.28  & 1451.499 & 5.3 \\
CASSINI & 2018-11-29 & 1452.338 & 1452.46  & 2.9 \\
CASSINI & 2018-12-04 & 1457.263 & 1457.537 & 6.6 \\
KUIPER  & 2018-12-09 & 1462.559 & 1462.892 & 8.0   \\
CASSINI & 2018-12-10 & 1463.248 & 1463.539 & 7.0   \\
CASSINI & 2018-12-11 & 1464.274 & 1464.545 & 6.5 \\
BOK     & 2018-12-17 & 1470.692 & 1470.811 & 2.9 \\
CAHA    & 2018-12-20 & 1473.355 & 1473.46  & 2.5
\enddata

\end{deluxetable*}

\begin{deluxetable*}{P{2cm}P{2cm}P{2cm}P{2cm}P{1cm}}[t]
\tablecaption{EIC-3 (2MUCD 20263) List of Observations} \label{tab:t3obslist}

\tablehead{\colhead{Telescope} & \colhead{Local Date} & \colhead{BJD Start ($-245700$)} & \colhead{BJD End ($-245700$)} & \colhead{Hours} } 

\startdata
KUIPER & 2018-12-09 & 1462.903 & 1463.049 & 3.5  \\
VATT   & 2018-12-18 & 1471.69  & 1471.944 & 6.1  \\
VATT   & 2018-12-19 & 1472.689 & 1473.051 & 8.7  \\
VATT   & 2018-12-20 & 1473.696 & 1473.955 & 6.2  \\
VATT   & 2018-12-28 & 1481.742 & 1482.059 & 7.6  \\
VATT   & 2018-12-29 & 1482.796 & 1483.064 & 6.4  \\
VATT   & 2018-12-30 & 1483.774 & 1484.064 & 7.0    \\
KUIPER & 2019-01-02 & 1486.675 & 1487.041 & 8.8  \\
KUIPER & 2019-01-03 & 1487.762 & 1488.039 & 6.6  \\
VATT   & 2019-01-08 & 1492.596 & 1493.044 & 10.8 \\
CAHA   & 2019-01-09 & 1493.326 & 1493.735 & 9.8  \\
CAHA   & 2019-01-10 & 1494.307 & 1494.726 & 10.1 \\
VATT   & 2019-01-10 & 1494.806 & 1494.985 & 4.3  \\
VATT   & 2019-01-11 & 1495.608 & 1496.034 & 10.2 \\
CAHA   & 2019-01-14 & 1498.27  & 1498.726 & 10.9 \\
CAHA   & 2019-01-15 & 1499.289 & 1499.717 & 10.3 \\
CAHA   & 2019-01-16 & 1500.281 & 1500.469 & 4.5  \\
LOT    & 2019-01-17 & 1500.962 & 1501.322 & 8.7  \\
LOT    & 2019-01-18 & 1502.043 & 1502.075 & 0.8  \\
VATT   & 2019-01-19 & 1503.589 & 1504.015 & 10.2 \\
VATT   & 2019-01-20 & 1504.586 & 1504.8   & 5.1  \\
VATT   & 2019-01-22 & 1506.657 & 1507.006 & 8.4  \\
VATT   & 2019-01-23 & 1507.588 & 1508.002 & 9.9  \\
BOK    & 2019-01-24 & 1508.605 & 1509.01  & 9.7  \\
BOK    & 2019-01-25 & 1509.582 & 1509.982 & 9.6  \\
VATT   & 2019-01-26 & 1510.591 & 1510.987 & 9.5  \\
VATT   & 2019-01-27 & 1511.582 & 1511.99  & 9.8  \\
LOT    & 2019-01-28 & 1511.959 & 1512.34  & 9.1  \\
VATT   & 2019-01-28 & 1512.61  & 1512.743 & 3.2  \\
LOT    & 2019-01-29 & 1512.952 & 1513.336 & 9.2  \\
LOT    & 2019-01-30 & 1513.968 & 1514.332 & 8.7  \\
VATT   & 2019-01-30 & 1514.679 & 1514.703 & 0.6  \\
LOT    & 2019-01-31 & 1514.973 & 1515.324 & 8.4  \\
VATT   & 2019-01-31 & 1515.663 & 1515.796 & 3.2  \\
LOT    & 2019-02-07 & 1521.973 & 1522.3   & 7.8  \\
LOT    & 2019-02-08 & 1523.045 & 1523.237 & 4.6  \\
LOT    & 2019-02-09 & 1524.031 & 1524.234 & 4.9  \\
LOT    & 2019-02-10 & 1524.997 & 1525.192 & 4.7  \\
LOT    & 2019-02-11 & 1526.023 & 1526.236 & 5.1  \\
LOT    & 2019-02-12 & 1526.967 & 1527.235 & 6.4  \\
LOT    & 2019-02-13 & 1528.016 & 1528.17  & 3.7  \\
LOT    & 2019-02-14 & 1528.994 & 1529.175 & 4.3 
\enddata

\end{deluxetable*}

\end{document}